# Dispersions of multi-walled carbon nanotubes in liquid crystals: new challenges to molecular theories of anisotropic soft matter

Longin N. Lisetski[a], Nikolai I. Lebovka[b]*, Sergei V. Naydenov[c], Marat S. Soskin[d]

[a]*Institute for Scintillation Materials, National Academy of Sciences of Ukraine, 60 Lenin Ave., 61001 Kharkov, Ukraine*

[b] *Institute of Biocolloidal Chemistry, NAS of Ukraine, 42 Vernadskii Prosp., Kyiv 03142, Ukraine*

[c] *Institute for Single Crystals, NAS of Ukraine, 60 Lenin Ave., 61001 Kharkov, Ukraine*

[d]*Institute of Physics, NAS of Ukraine, 46 Prospect Nauki, Kyiv 03650, Ukraine*

## ABSTRACT

Aggregation of carbon nanotubes dispersed in nematic liquid crystalline medium is discussed. A model is proposed, which assumes that the aggregates consist of a "skeleton" formed by stochastically arranged nanotubes and a "shell" ("coat") of incorporated and adjacent nematic molecules. The aggregates of this type can be considered as large quasi-macroscopic particles in the nematic matrix. The resulting composite system represents a new type of complex molecular liquids involving self-organization of particles in anisotropic medium. Many essential features and implications of the theoretical model (e.g., effects of concentration of the nanotubes, their aspect ratio and orientational order parameter on the size and fractal dimensionality of the aggregates formed, as well as on the rate of aggregation) are in good agreement with experimental data obtained by various physical methods.

---

*Corresponding author. Institute of Biocolloidal Chemistry, NAS of Ukraine, 42 Vernadskii Prosp., Kyiv 03142, Ukraine.  Tel.: +38044 424 0378; fax: 380 44 424 0378.
*E-mail address*: lebovka@gmail.com (Nikolai Lebovka)

# 1. Introduction

Recently, great attention has been attracted by the anisotropic soft matter systems including particles of different size, shape and aspect ratio. Among the most interesting are multi-walled carbon nanotubes (NT) dispersed in liquid crystalline (LC) media [1-3]. These composites display many interesting and unexpected features presenting both prospects of practical application and challenges to molecular theories of anisotropic liquids.

For theoretical purposes, two different types of LC+NT dispersions can be assumed. At low NT concentrations, each NT aligns several molecular layers of nematic molecules on a short-range scale [4,5]. On the long-range scale, nematic as an anisotropic fluid orients the NTs that can be formally considered as non-mesogenic dopants [6-8]. At higher NT concentrations (above a certain "percolation threshold" [9]), NTs show a strong tendency for aggregation in a LC medium with formation of random network of contacting aggregates. In most cases, LC+NT dispersions obtained by conventional ultrasonication procedures contain a number of NT "bundles", i.e., more or less large agglomerations of nanotubes that are strongly anchored to each other in the lateral direction. These bundles are often clearly visible in an optical microscope, and in principle they can be mechanically removed for making the dispersion more homogeneous. Such "quasi-homogeneous" dispersions with minimized quantity and size of bundles can also be obtained by careful sonication under certain optimized conditions. However, immediately after such sonication spontaneous processes of NT aggregation begin. The spanning networks of nanotubes arise and grow with time due to incorporation of individual NTs from the quasi-homogeneous dispersion. This process of aggregation manifests itself in changes of certain macroscopic properties of LC+NT dispersions (e.g., optical transmission, electrical conductivity, singular and polarization structure of transmitted laser beam, etc.), which can be measured on the time scale of several hours [9]. An important point is that such "secondary" aggregates formed in LC+NT dispersions are substantially different from the initial NT bundles occurring in imperfectly sonicated samples. These aggregates are rather "loose" and are obviously of fractal nature. The preliminary study of such aggregates was reported in our recent paper [10]. It was shown that the ramified aggregates of nanotubes capture (incorporate) surrounding LC molecules, and their volume becomes by 2-2.5 orders greater than the total volume of the NTs involved. Such aggregates can be considered as large quasi-macroscopic particles in the nematic LC matrix. The supramolecular formations of this type were named as "S-aggregates" [10]. It was shown by methods of singular optics that anisotropic micro-sized LC cladding of these ramified aggregates initiates strong speckled scattering with induced optical singularities [11].

Theoretical studies of LC+NT dispersions are not numerous [12-14], and they were related, in fact, only to the "idealized" dispersions, with no account for the aggregation processes. This work, assuming that the above-mentioned LC+NT dispersions containing *C*-aggregates can be considered as a new type of complex liquids, is aimed at development of a preliminary theoretical approach to description of such anisotropic systems.

## 2. Theory

*2.1. Formation of fractal aggregates in dispersions of carbon nanotubes in a liquid crystalline medium*

The starting point of our theoretical consideration is a tentative picture of the fractal clusters of rods formed in colloidal rod suspensions as presented by Solomon and Spicer [15] (more specifically, their arrangement is shown in Fig.3b of their paper). Fig.1 presents a simplified picture of such structure adjusted for the specific case of nanotubes forming loose fractal aggregates in the anisotropic liquid crystal medium.

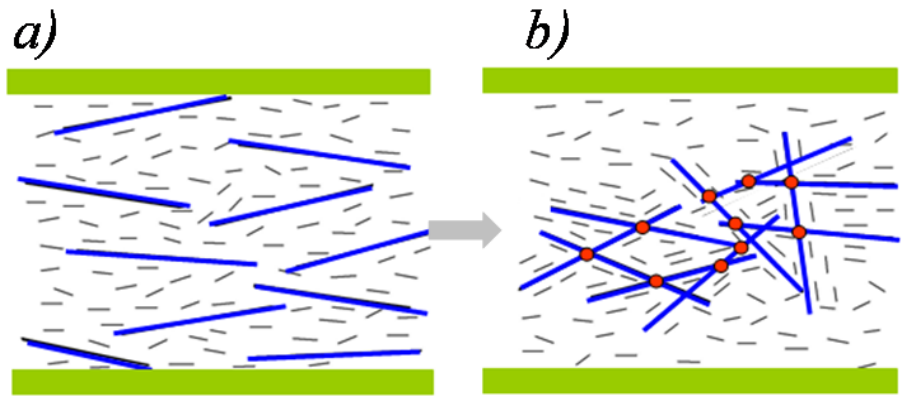

Fig.1. A simplified scheme of aggregation of carbon nanotubes dispersed in nematic liquid crystalline medium. The nanotubes are initially dispersed homogeneously after intensive ultrasonication (a) and then appear the ramified aggregates of fractal nature with a large number of nematic molecules incorporated into the "skeleton" of nanotubes (b).

Let us consider the nanotubes (NT) comprising S-aggregates as linear elements forming a compact skeleton that bounds a confined space. We determine the volume encompassed by such skeleton using only one characteristic spatial dimension of NT – its linear length $l$, which is fully justified if the aspect ratio $r \gg 1$ ($r = l/d$, where $d$ is the characteristic lateral dimension (NT diameter)).

Under considerations of geometrical similarity, the volume of a cluster with all its links having the same characteristic linear dimension $l$ can be presented as $V_{cl} = V_{cl}(l) = A_{cl} l^3$, where $A_{cl}$ is a certain constant depending on the cluster geometry (e.g., $A_{cl} = 1$ for cubic structure, $A_{cl} =$

$\sqrt{2}/12$ for spatial structure formed by tetrahedrons, etc.). Similarly, the total volume of the nanotubes forming the skeleton is $V_{nt}(l) = B_{nt}\, l\, g(d)$, where $B_{nt}$ is a certain new constant. The value of $B_{nt}$ depends upon cluster geometry. Function $g(\ldots)$ determines the volume of a single nanotube in the cluster as function of its characteristic cross-section area. The form of $g(\ldots)$ in the case of a regular skeleton structure depends on geometry of isolated nanotubes, accounting for the bends, fractures, non-cylindrical shape, scatter of lateral dimensions, etc. In the case of fractal geometry of the cluster, $g(\ldots)$ should "feel" how the nanotubes are interconnected. Then instead of $g(x) \sim x^2$ one can choose a characteristic self-similar dependence of a typical area of the irregular surface, $g_f(x) \sim x^{d_f}$, where $d_f$ is the fractal dimension of the cluster surface. For regular geometry $d_f = 2$; for fractal geometry, the value of $d_f$ is fractional.

For regular skeleton geometry, the total volume of the constituent nanotubes is $V_{nt}(l) = B_{nt} l g(d)$. Accounting that $g(d) = (\pi/4)d^2$ for cylindrical nanotubes, $V_{nt}(l) = B_{nt}(\pi/4)\, l^3 r^{-2}$. The coefficient $B_{nt}$ depends on the type of the regular skeleton, e.g., $B_{nt} = 3\pi/2$ for tetrahedral skeleton, $B_{nt} = 3\pi$ for cubic cells, etc.

Thus, the volume ratio of the cluster and the regular skeleton (i.e., the ratio of the volume encompassed by the skeleton of nanotubes and the total volume of nanotubes involved) is $W_{aggr} = V_{cl}/V_{nt} = (A_{cl}/B_{nt})r^2 = Ar^2$. With increase of aspect ratio (i.e., with longer or thinner nanotubes), $W_{aggr}$ is growing rapidly, while the volume with nematic molecules "captured" by the skeleton of nanotubes (the volume of the "coat") increases much slower (not faster than linearly). The coefficient $A$ is determined by the specific geometry of the skeleton, and its values can vary within $10^{-1} > A > 10^{-2}$ ($A = 1/3\pi \sim 0.1$ for cubic cells, $A \sim 0.025$ for tetrahedral cells).

Thus, $W_{aggr}$ is the ratio of the volume of an S-aggregate (i.e., of a loose skeleton formed by nanotubes together with "captured" molecules of the dispersion medium (nematic LC) both "inside" the skeleton and in the nearest coordination layers around it) and the total volume of the nanotubes forming the skeleton. If $c = V_{nt}/V$ is the volume fraction (or mass fraction, if the difference between the densities of the dispersion medium and the nanotubes can be neglected) of nanotubes that were introduced into the matrix, $C_{aggr} = cW_{aggr}$ can be considered as effective volume concentration of S-aggregates in the matrix (solvent, dispersion medium) where these aggregates are formed.

In the case of fractal geometry, both the surface of the skeleton formed by the nanotubes and the outer surface of S-aggregate formed by the aggregated nanotubes and captured molecules of the dispersion phase will be of fractal character. Accounting for the fact that S-aggregates in the orientationally ordered matrix are formed by orientationally ordered nanotubes, it is reasonable to expect that S-aggregates are essentially oblate (flattened). This leads us to a heuristic formula

$$W_{aggr} = \frac{V_{cl}}{V_{nt}} = Ar^{d_f} \qquad (1)$$

Correspondingly, the effective concentration of aggregates in the LC+NT dispersion is

$$C_{aggr} = cW_{aggr} = c\frac{A_{cl}}{B_{nt}}r^{d_f} = cAr^{d_f} \qquad (2)$$

Thus, the effective concentration of S-aggregates at a given initial concentration of nanotubes appears to be a universal function of the aspect ratio, which is rather attractive. The proportionality coefficient $A$ is determined by specific geometries of the nanotubes and the fractal skeleton structure formed by them; it can be considered as constant for a given type of the nanotube dispersions. As noted above, its value can be taken within $0.025 - 0.1$.

This formula accounts, at semi-empirical level, for all main factors affecting the observed results of the nanotube aggregation – the initial concentration of nanotubes $c$, their aspect ratio $r$, geometry of cluster formation $A$ and fractal dimensionality of the formed aggregates $d_f$. This formula can be easily verified experimentally: thus, for two different NT+LC dispersions prepared and studied under the same conditions, but differing in certain parameters (e.g., aspect ratio $r$ and/or nanotube concentration $c$), the other measured characteristics should be related by (2).

Preliminary results of such verification can be obtained using data from [9]. Fig.3 of this paper shows $d_f$ values for three concentrations of nanotubes (generally speaking, for three specific cases of S-aggregate formation). According to (1.3), at $r = 100$ and $A = 0.1$ we get, for $c = 0.1\%$ and $d_f = 1.81$, the value of $W_{eff} = 417$, i.e., the total volume of the formed S-aggregates with fractal dimensionality 1.81 will be $C_{aggr} = 41.7\%$ with respect to the total volume of the system (the remaining 58.3% is the fraction of volume occupied by orientationally ordered nematic molecules that remain "free", i.e., not captured by the aggregates). Accordingly, for $c = 0.01\%$ and $d_f = 1.54$ we get $C_{aggr} = 12\%$. This is in a semi-quantitative agreement with the results of microscopic observations: the aggregates are relatively few and visually occupy ~ 10% of the total area in dispersions with 0.01% NT, while in dispersions with 0.1% NT there are much more aggregates, their size is larger, and they occupy about one half of the visible area.

A similar effect is experimentally observed when we do not increase the NT concentration, but just time changes occur in a freshly dispersed sample (incubation). In several hours the fraction of area occupied by the aggregates substantially increases ([9], Fig.6).

*2.2. NT aggregates in LC medium: characteristics of the aggregate formation process*

As a further step, we will consider in more detail the process of formation of S-aggregates in the liquid crystalline medium. The model scheme of structural arrangement of NTs in the orientationally ordered (nematic) LC is assumed, which was earlier proposed [16] for explaining of the anomalous increase of electrical conductivity with concentration in nematic NT dispersions (Fig.2).

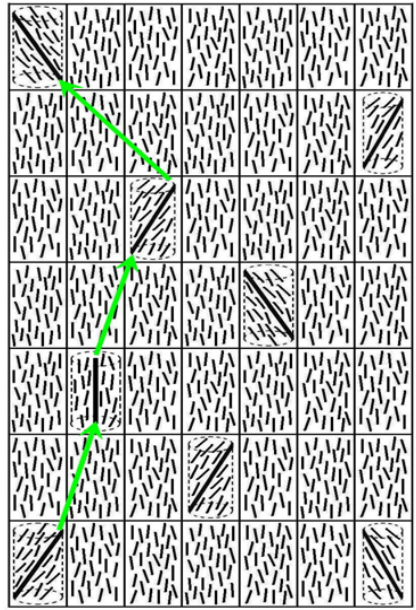

Fig.2. A model picture describing "quasi-homogeneous" distribution of isolated carbon nanotubes in nematic liquid crystalline medium. The path for electrical transport is shown.

It can be noted that the imaginary "cell" of such structured pattern is rather similar to the picture of a nanotube with adjacent layers of nematic molecules presented in [4,5]. Only one "quasi-lamellar" layer is considered for simplicity. If the volume concentration of the nanotubes is $c$, which is also their concentration in the quasi-lamellar layer, then NT concentration in the planar projection ("linear concentration") is $c^{1/2}$. If $c = 0.001$ (i.e., 0.1%), $c^{1/2} = 0.033$ (~3%). At "linear concentration" $c^{1/2}$, the average distance between the nanotubes in the quasi-lamellar layer is $m = dc^{1/2}$ =5 nm x 1/0.033, i.e., ~ 150 nm. Here, the typical dimensions of the used multi-walled carbon nanotubes, $l = 500$ nm (length) and $d =5$ nm (diameter) are taken into account.

In a nematic phase with long-range orientational order parameter $S < 1$, long axes of nanotubes deviate from the preferred direction by a certain angle $\varphi$. The scatter of $\varphi$ values is determined by the orientational order parameter

$$S = \langle 3\cos^2 \varphi - 1 \rangle / 2 \qquad (3)$$

It may be assumed that the system is time stable ("vitrificated") with parameters having statistically averaged values. Then the average distance between the nanotubes in the quasi-lamellar layer may be estimated as $m = l\sin\varphi$. On the other side, $m = dc^{1/2}$ (see above). Then, taking not fully rigorous but rather adequate relationship $l\sin\varphi = dc^{1/2}$ and introducing the "aspect ratio" of the nanotubes $r = l/d$, we obtain

$$\sin\varphi = \frac{d}{lc^{1/2}} = \frac{1}{rc^{1/2}} \tag{4}$$

At $c = 0.001$ (0.1%) and $l = 500$ nm, $d = 5$ nm, we get $\sin\varphi \sim 0.3$. Thus, $\varphi \sim 17°$, $\sin^2\varphi \sim 0.09$, $\cos^2\varphi = 1 - \sin^2\varphi \approx 0.91$, and $(3\cos^2\varphi - 1)/2 = 0.865$. I.e., the estimated orientational order parameter of the nanotubes $S$ is of order of 0.865 and it is close to the experimental estimates of this value. At $c = 0.0007$ (~0.07%) that corresponds to the "percolation threshold" for formation of aggregates network [11], the similar numerical estimations give $S \sim 0.796$.

Thus, the value of $d/lc^{1/2}$ (or $1/rc^{1/2}$) can be a certain criterion for commencement of the intensive formation of aggregates in the quasi-homogeneous LC+NT dispersion (at a given value of the orientational order parameter $S$).

If $S$ substantially decreases, the aggregation threshold should also decrease. Recent experimental data [17] show that NT dispersions in the nematic E7 (liquid crystal mixture on the basis of cyanobiphenyls) are more stable than in 5CB (4-pentyl-4'-cyanobiphenyl). This can be easily explained accounting for much higher nematic-isotropic transition temperature of E7 as compared with 5CB. So, under the same conditions (i.e. at the room temperature) the value of order parameter $S$ is higher in E7 than in 5CB. Experimental data [18] evidence, that aggregation is much more intensive in the isotropic than in nematic phase. The effects of short-range orientational ordering can influence the aggregation of NTs. However, these effects noticeably reduce in the isotropic phase, where the effective order parameter $S$, is equal to zero.

*2.3. Kinetics of aggregation*

The above considerations refer to a "static" picture. The first approximation in description of dynamics (kinetics) of formation of the aggregates can be as follows.

If we consider an arbitrary function reasonably describing the time dependence of the aggregation, we can assume that it can be expanded in series over a certain parameter, and the first terms of this expansion can be assumed as linear or quadratic. If $c_0$ is the critical concentration required for formation of the aggregates in the static model and the actual concentration of NTs in the initial (freshly prepared) quasi-homogeneous dispersion is $c$, the rate of aggregation can be written as $v_{aggr} = v_0 + k(c - c_0)$. If other system parameters remain unchanged (i.e., the same nanotubes are dispersed in the same nematic matrix), lower critical concentration $c_0$ (determined from the static model) leads to higher rate of aggregation. Lower values of $S$ (implying smaller $c_0$) will result in faster aggregation, i.e., the same degree of aggregation will be achieved much faster. Thus, the product $rc^{1/2}$ can be considered as a certain

"similarity criterion" that determines the value of NT concentration ensuring approximately similar conditions of aggregation for a given aspect ratio (or vice versa).

Another approximation takes into account that aggregates of nanotubes as shown in Fig.1 can be considered, according to [15], as fractal clusters of rods formed by diffusion-limited cluster aggregation. The simple Smoluchowski approach was applied for description of the Brownian motion of individual nanotubes in order to estimate the half-time of aggregation. In our approach, the multi-walled carbon nanotubes (NTs) were simulated by cylindrical particles with huge aspect ratio $r=l/d (r>>1)$, where $l$ and $d$ are the nanotube length and diameter, respectively. Under assumption that cylindrical particles can aggregate when the distance between them is of the order of their length $l$, the half-time of aggregation $\theta$ can be estimated as

$$\theta=(4D\pi nl)^{-1}, \quad (5)$$

where $D$ is the diffusion coefficient and $n$ is the numerical concentration of the particles. These values can be expressed as $D=kT\ln r/(3\pi\eta rd)$ and $n=(\rho/\rho_c)C/V_c$, respectively, where $kT$ is the thermal energy; $\eta$ is viscosity, $\rho/\rho_c$ is the ratio of densities of 5CB and the nanotubes ($\rho/\rho_c\approx 0.5$), $C$ is the mass fraction of NTs in 5CB, and $V_c=\pi d^3 r/4$ is the volume of a single nanotube.

Finally, the following relation can be obtained:

$$\theta \approx 3\pi d^3 \eta/(8kTC)\,(r/\ln r). \quad (6)$$

Taking into account that viscosity of 5CB $\eta \approx 0.1$ Pa·s, $d = 20$ nm, $r = 500$, temperature $T = 300$ K, we obtain $\theta \approx 180$ s for $C = 0.01$ % wt and $\theta \approx 18$ s for $C = 0.1$ % wt.

Further steps in development of a general theoretical description of the process of NT aggregation in a liquid crystalline medium should probably involve a certain combination of the "orientational" and "diffusion" approaches. In the latter case, high anisotropy of viscosity in the nematic phase, especially in the vicinity of nanotubes, should be taken into account. Another point is that formation of NT aggregates itself can be a complicated process involving several stages.

The carried out experiments with dispersions of multi-walled carbon NTs in 5CB show that the initial stage of aggregation results in formation of the ramified, or loose, aggregates [19-21]. The structure of aggregates and physical characteristics of the LC+NT composites get stabilized only several hours or days after ultrasonication. Such behaviour can be explained by the process of compaction of the ramified, or loose, aggregates ("L-aggregates") formed in the "fresh" suspension into more dense aggregates ("C-aggregates") occurring in the thermally incubated composite samples. However, the mechanism of compacting is still unclear, and understanding of further details of the aggregation process, including specific features of orientational and translational diffusion motions, as well as inter-particle interactions in a LC medium, is required. It should stimulate the future experimental and theoretical studies of LC+NT systems.

## 3. Conclusions

The theoretical model is proposed as a basis for understanding of a large set of experimental data on dynamics of NT aggregate formation in dispersions of multi-walled carbon nanotubes in nematic liquid crystal 5CB that were obtained using different techniques: (a) microscopic observation of microstructure, (b) study of light transmission changes in the point of nematic-isotropic phase transition ($T_i$), (c) differential scanning calorimetry in the vicinity of $T_i$, (d) measurements of the temperature and concentration dependences of electrical conductivity, including conductivity vs. applied voltage behavior in the Freedericks transition geometry, (e) measurement of birefringent structure of LC cladding around NT aggregates and induced optical singularities [10, 19-21].

The process of NT aggregation can be generally described in the following way. The individual nanotubes homogeneously dispersed in the nematic matrix after ultrasonication gradually assemble into fractal aggregates incorporating 5CB molecules into the "micropores" of an aggregate "skeleton" produced at loose arrangement of NTs. For example, at 0.1% wt of NTs the aggregates arising after several hours or days of incubation behave like micron-sized small particles occupying up to 20-40% of the bulk nematic volume. Higher concentration and higher aspect ratio of NTs strongly favor the aggregation without changing the general picture of the process.

## Acknowledgement

This work was supported in part by Projects 2.16.1.4, 2.16.1.7, ISTCU Project 4687.